\documentstyle[sprocl,psfig]{article}
\def\Journal#1#2#3#4{{#1} {\bf #2}, #3 (#4)}

\def\apjl{{\em Astrophys. J. Lett.}}
\def\apjs{{\em Astrophys. J. Suppl.}}
\def\mnras{{\em Mon. Not. Roy. Astr. Soc.}}
\def\nat{{\em Nature}}

\def\PLB{{\em Phys. Lett.}  B}

\def\etal{et\ al.}
\def\CIV{C~{\sc iv}}

\def\hMpc{$h^{-1}$~Mpc}
\def\gtrsim{\mathrel{\hbox{\rlap{\hbox{\lower4pt\hbox{$\sim$}}}\hbox{$>$}}}}

\begin{document}
\title{LARGE--SCALE STRUCTURE AS SEEN FROM QSO
ABSORPTION--LINE SYSTEMS}
\author{J.M. QUASHNOCK, D.E. VANDEN BERK, D.G. YORK}
\address{Department of Astronomy and Astrophysics, University of Chicago,\\
5640 S. Ellis Avenue, Chicago, IL 60637, USA}

\maketitle\abstracts{
We study clustering on very large scales ---
from several tens to hundreds of comoving Mpc ---
using an extensive catalog of heavy--element QSO absorption--line systems.
We find significant evidence that \CIV\ absorbers are clustered on
comoving scales of 100 \hMpc\ ($q_0=0.5$) and less.
The superclustering is present even at high redshift ($z\sim 3$);
furthermore, it does not appear that the superclustering scale (comoving)
has changed significantly since then.  Our estimate of that scale
increases to 240 \hMpc\ if $q_0=0.1$, which is larger
than the largest scales of clustering seen at the present epoch.
This may be indicative of a larger value of $q_0$, and hence $\Omega_0$.
We identify 7 high--redshift supercluster candidates,
with 2 at redshift $z\sim 2.8$.
The evolution of the correlation function
on 50 \hMpc\ scales is consistent
with that expected in cosmologies with
$\Omega_0= $ ranging from 0.1 to 1.
Finally, we find no evidence for clustering on scales greater than 100 \hMpc\
($q_0=0.5$) or 240 \hMpc\ ($q_0=0.1$).}

It has been recognized for some time now that
QSO absorption line systems
are particularly effective probes of large--scale structure
in the universe.\,\cite{CMY85}
This is because the absorbers trace matter lying on the QSO line of sight,
which can extend over a sizable redshift interval out to high redshifts.
Thus, the absorbers trace both the large--scale structure
and its evolution in time,
since the clustering pattern can be examined as a function of
redshift out to $z\sim 4$.
The evolution of large--scale structure is of great interest, since,
in the gravitational instability picture,
it depends sensitively on 
$\Omega_0$.\,\cite{Peebles93}

Here we study clustering by computing
line--of--sight correlations of \CIV\ absorption line systems,
using a new and extensive catalog of absorbers.\,\cite{Y97}
(A more complete version of this work has appeared elsewhere.\,\cite{Quash96})
This catalog contains data on all QSO
heavy--element absorption lines in the literature.
It is an updated version of
the York \etal\ (1991) catalog,\,\cite{Y91}
but is more than twice the size, with over 2200
absorbers listed over 500 QSOs, 
and is the largest sample of heavy--element absorbers compiled to date. 

Figure~1 shows the \CIV\ line--of--sight correlation function, 
$\xi_{\rm aa}$, as a function of absorber comoving
separation, $\Delta r$,
for the entire sample of absorbers.
The results are shown for both a $q_0=0.5$ ({\it left panel}, 25 \hMpc\ bins)
and a $q_0=0.1$ ({\it right panel}, 60 \hMpc\ bins) cosmology.\footnote
{Larger bins are required for $q_0=0.1$ because, at high redshift, a larger
comoving separation $\Delta r$ arises
from a fixed redshift interval $\Delta z$.}
The vertical error bars through the data points are $1\, \sigma$ errors in the
estimator for $\xi_{\rm aa}$,
which differ from the $1\, \sigma$ region of scatter ({\it dashed line},
calculated by Monte Carlo simulations)
around the no--clustering null hypothesis.

Remarkably, there appears to be significant clustering in
the first four bins of Figure~1: 
The positive correlation seen in the first four bins of
Figure~1 has a significance of $5.0\, \sigma\ $.
Therefore, there is significant evidence of clustering of matter
traced by \CIV\ absorbers on scales up to 100 \hMpc\ ($q_0=0.5$)
or 240 \hMpc\  ($q_0=0.1$).
There is no evidence from Figure~1
for clustering on comoving scales greater than these.

\begin{figure}
\psfig{figure=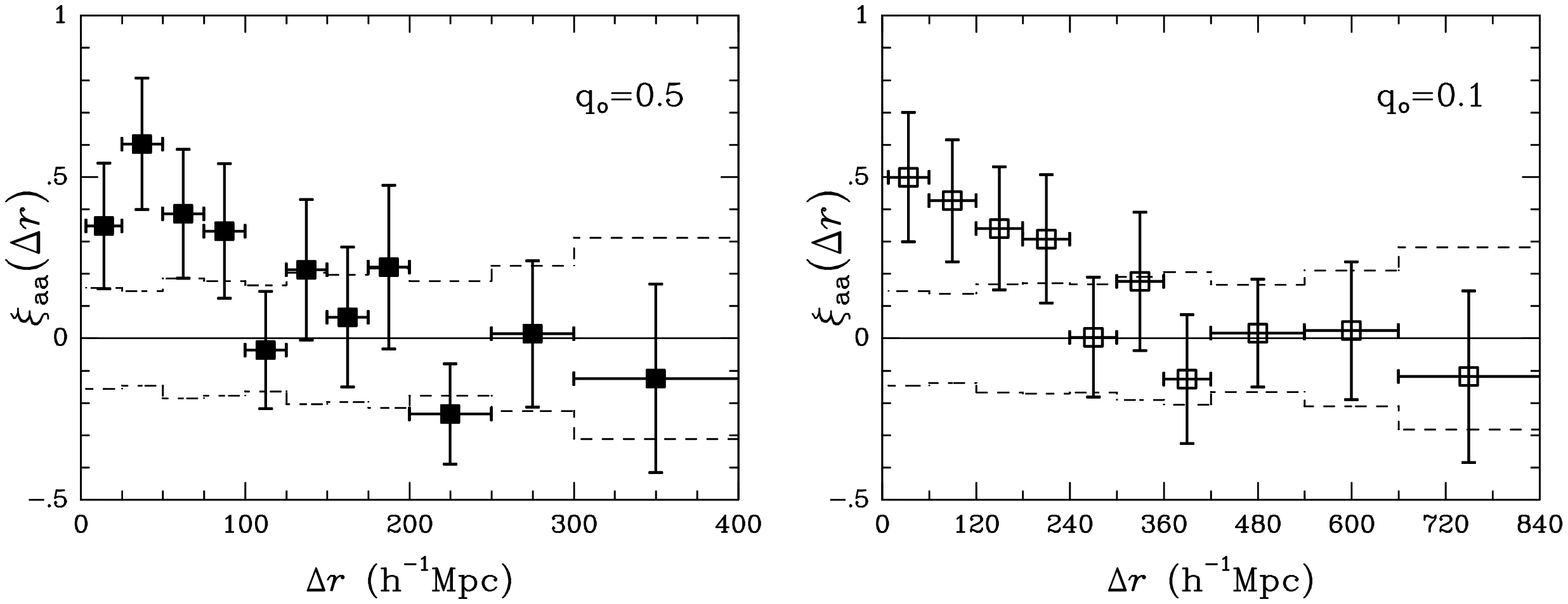,width=4.7in,angle=-90} \caption{Line--of--sight 
correlation function of \CIV\ absorbers as a function of
absorber comoving separation (from Ref. 4,
\copyright 1996 by The American Astronomical Society).\label{fig1}}
\end{figure}

We have investigated the evolution of the superclustering by dividing
the absorber sample into three approximately equal redshift sub--samples;
namely, low ($1.2<z<2.0$), medium ($2.0<z<2.8$), 
and high ($2.8<z<4.5$) redshift.
We find that the significant superclustering seen in Figure~1 is
present in all three  redshift sub--samples,
so that the superclustering  is present
even at redshift $z \gtrsim 3$.
Furthermore, it does not appear that the superclustering scale,
in comoving coordinates,
has changed significantly since then.

We have examined the clustering signal more closely
and find that a large portion comes from
7 QSO lines of sight that have groups of
4 or more \CIV\ absorbers
within a 100 \hMpc\ interval ($q_0=0.5$).
(From Monte Carlo simulations,
we expect only $2.7\pm 1.5$ QSOs with such groups.)
We have found two potential superclusters, at redshift $z\sim 2.8$,
among these groups.

The superclustering is
indicative of generic large--scale clustering
in the universe, out to high redshift $z \gtrsim 3$,
on a scale frozen in comoving coordinates
that is --- if $q_0=0.5$ --- similar to the size of the voids and walls in
galaxy redshift surveys of the local 
universe.\,\cite{Kir81}$^{-}$\,\cite{Landy96}
It also appears consistent with the general finding \cite{Broad90,Ein97}
that galaxies are clustered in a regular pattern on very large scales,
although we have not confirmed that  there is
quasi--periodic clustering with power
peaked at $\sim$ 128 \hMpc .

Our estimate of the superclustering scale increases to 240 \hMpc\ if
$q_0=0.1$ (see Figure~1),
which is larger than the largest scales of clustering known at present.
If the structures traced by
\CIV\ absorbers are of the same nature as those seen locally
in galaxy redshift surveys,
the superclustering scale should have
a value closer to 100 \hMpc\ .
This may be indicative of a larger value of $q_0$,  and hence $\Omega_0$.

We find that the evolution of the correlation function
on 50 \hMpc\ scales is consistent
with that expected in cosmologies with
density parameter ranging from $\Omega_0= $ 0.1 to 1.

\section*{Acknowledgments}
JMQ is supported by the Compton Fellowship -- NASA grant GDP93-08.
DEVB was supported in part by the Adler Fellowship at the University of
Chicago, and by NASA Space Telescope grant GO-06007.01-94A.

\end{document}